\documentclass[a4paper,twocolumn,showpacs,superscriptaddress,floatfix,nofootinbib]{revtex4-1}

\usepackage{latexsym}
\usepackage{amssymb}
\usepackage{amsfonts}
\usepackage{amsmath}
\usepackage{bm}
\usepackage{graphicx}   
\usepackage{color}
\usepackage{subfigure}
\usepackage{times}
\usepackage{units}
\usepackage{hyperref}

\begin{document}

\title[Binary neutron star mergers]{Binary neutron star mergers after GW170817} 

\author{Riccardo Ciolfi}
\address{INAF, Osservatorio Astronomico di Padova, Vicolo dell'Osservatorio 5, I-35122 Padova, Italy}
\address{INFN, Sezione di Padova, Via Francesco Marzolo 8, I-35131 Padova, Italy}

\date{\today}

\begin{abstract}
\noindent The first combined detection of gravitational waves and electromagnetic signals from a binary neutron star (BNS) merger in August 2017 (event named GW170817) represents a major landmark for the ongoing investigation on these extraordinary systems. 
In this short review, we introduce BNS mergers as events of the utmost importance for astrophysics and fundamental physics and discuss the main discoveries enabled by this first multimessenger observation, which include compelling evidence that such mergers produce a copious amount of heavy r-process elements and can power short gamma-ray bursts. We further discuss key open questions left behind on this event and BNS mergers in general, focussing the attention on the current status and limitations of theoretical models and numerical simulations.  
\end{abstract}

\maketitle

\section{Introduction}
\label{intro}

\noindent  Binary neutron star (BNS) mergers are among the most intriguing events known in the Universe, characterized by an impressive scientific potential spanning many different research fields in physics and astrophysics.
Investigating them offers a unique opportunity to understand hadronic interactions at supranuclear densities and the equation of state (EOS) of matter in such extreme conditions and, at the same time, to gain crucial insights on the strong gravity regime, high energy astrophysical phenomena of primary importance like gamma-ray bursts (GRBs), the origin of heavy elements in the local Universe, formation channels of compact object binaries, and cosmology (e.g., \cite{Faber2012,Baiotti2017} and refs. therein). 

The merger of two neutron stars (NSs) is accompanied by a strong emission of gravitational waves (GWs) and by a rich variety of electromagnetic (EM) signals covering the entire spectrum, from gamma-rays to radio. 
Such a unique combination of signals makes these systems ideal multimessenger sources and allows us to observe them up to cosmological distances.
Moreover, among their EM ``counterparts'', BNS mergers have long been thought to be  responsible for short GRBs (SGRBs) 
\cite{Paczynski1986,Eichler1989,Narayan1992,Barthelmy2005a,Fox2005,Gehrels2005,Berger2014} as well as radioactively-powered ``kilonova'' transients associated with r-process nucleosynthesis of heavy elements \cite{Li1998,Rosswog2005,Metzger2010}.\footnote{Merging mixed binaries composed by a NS and a back hole (BH) share most of the above features, being also promising GW sources, potential SGRB central engines, and potential sources of radioactively-powered kilonovae. 
However, the properties of the emitted signals might be very different. Here, we focus on BNS mergers only and refer the reader to other reviews for the case of NS-BH binary mergers (e.g., \cite{Shibata2011LRR}).} 

A major step forward in the study of BNS mergers was made possible by the first GW detection of this type of events by the LIGO and Virgo Collaboration in August 2017 (event named GW170817) \cite{LVC-BNS}. This merger was also observed in the EM spectrum, via a collection of gamma-ray, X-ray, UV, optical, IR, and radio signals, thus providing as well the first multimessenger observation of a GW source \cite{LVC-MMA}.
Such a breakthrough led to a number of key discoveries, including a striking confirmation that BNS mergers can launch SGRB jets \cite{LVC-GRB,Goldstein2017,Savchenko2017,Troja2017,Margutti2017,Hallinan2017,Alexander2017,Mooley2018a,Lazzati2018,Lyman2018,Alexander2018,Mooley2018b,Ghirlanda2019} and are ideal sites for r-process nucleosynthesis (e.g., \cite{Arcavi2017,Coulter2017,Pian2017,Smartt2017,Kasen2017}; see also \cite{Metzger2019LRR} and refs.~therein), the first GW-based constraints on the NS EOS \cite{LVC-170817properties} and the Hubble constant \cite{LVC-Hubble}, and more. 
The most important lessons learned from this event are discussed in the next Section (\ref{170817}). 

\begin{figure*}[t!]
\begin{center}
\includegraphics[width=1.0\textwidth]{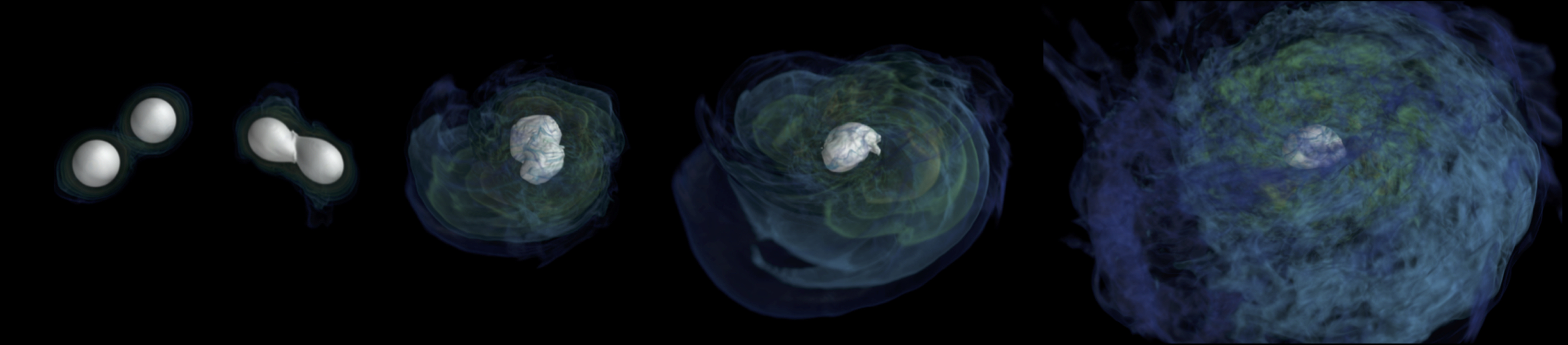}
\end{center}
\caption{Example of BNS merger simulation in GRMHD (from the models presented in \cite{Ciolfi2017}). The temporal sequence shows the bulk of the NS(s) in white together with color-coded isodensity surfaces.}
\label{fig1}
\end{figure*}

Together with the remarkable results mentioned above, the GW170817 event also left behind a number of open questions, in part related to details of the merging process that remained only poorly constrained. 
For instance, the remnant object resulting from the merger appears to be most likely a metastable massive NS that eventually collapsed into a BH, but the lack of clear indications on its survival time until collapse leaves doubts on the nature of the SGRB central engine, which could have been either the massive NS or the accreting BH (see, e.g., \cite{Ciolfi2018} for a recent review).
Theoretical modelling of the merger process via general relativistic magnetohydrodynamics (GRMHD) simulations (see Fig.~\ref{fig1}) offers the best chance to tackle the open issues and to establish a reliable connection between the merger and post-merger dynamics and the observable GW and EM emission
(e.g., \cite{Ciolfi2020b} and refs.~therein).
In Section~\ref{open}, we briefly report on the status of the investigation in this direction, with reference to specific challenges posed by the GW170817 event. 
Finally, concluding remarks are given in Section~\ref{concl}.

\section{The BNS merger of August 2017}
\label{170817}

\noindent The characteristic ``chirp'' signal of GW170817, with both frequency and amplitude increasing in time up to a maximum, leaves no doubts that the source was a merging compact binary with component masses fully consistent with being NSs \cite{LVC-BNS}. In addition, the BNS nature of the source is arguably reinforced by the EM counterparts observed along with GWs \cite{LVC-MMA}. Under the BNS assumption, this single detection significantly improved our estimate for the corresponding local coalescence rate (the value reported in \cite{LVC-BNS} being $R=1540^{+3200}_{-1220}$\,Gpc$^{-3}$yr$^{-1}$).\footnote{Under the assumption that GW190425 was also a BNS merger, the updated rate would be $R=250-2810$\,Gpc$^{-3}$yr$^{-1}$ \cite{LVC-190425}.}

For this event, most of the information inferred from GWs came from the inspiral phase up to merger, while the lower detector sensitivity at frequencies above 1\,kHz did not allow for a confident detection of the post-merger signal \cite{LVC-BNS}.
Despite such limitation, it was possible to start placing the first limits on the NS tidal deformability and thus to constrain the range of NS EOS compatible with the event (e.g., \cite{LVC-170817properties}; see also \cite{Kastaun2019}), by measuring finite-size effects (i.e.~deviations from the point-mass waveform) in the last orbits of the inspiral.
Moreover, combining the luminosity distance derived from GWs with the EM redshift measurement allowed by the identification of the host galaxy (NGC\,4993), it was possible to obtain the first constraints on the Hubble constant based on a GW standard siren determination \cite{LVC-Hubble}.

The observation of a gamma-ray signal emerging about 1.74\,s after the estimated time of merger enabled us to confirm that GWs propagate at the speed of light with a precision better than $10^{-14}$ \cite{LVC-GRB}, which excluded a whole range of gravitational theories beyond general relativity.
At the same time, this high energy signal (named GRB\,170817A) was found potentially consistent with a SGRB, although orders of magnitude less energetic than any other known SGRB \cite{LVC-GRB}.
Combining the prompt gamma-ray emission with the multiwavelength afterglows (in X-ray, optical, and radio) monitored for several months, it was possible to eventually converge to the following picture \cite{LVC-GRB,Goldstein2017,Savchenko2017,Troja2017,Margutti2017,Hallinan2017,Alexander2017,Mooley2018a,Lazzati2018,Lyman2018,Alexander2018,Mooley2018b,Ghirlanda2019}:
(i) the merger remnant launched a highly relativistic jet (Lorentz factor $>\!10$), in agreement with the consolidated GRB paradigm  (e.g., \cite{piran2004,Kumar2015}); (ii) the burst was observed off-axis by $15^\circ\!-\!30^\circ$ and the low energy gamma-ray signal detected was not produced by the jet core, but rather by a mildly relativistic outflow moving along the line of sight; (iii) the on-axis observer would have seen a burst energetically consistent with the other known SGRBs.
This provided the long-awaited compelling evidence that {\it BNS mergers can generate SGRBs}. 
Furthermore, the off-axis view of a nearby ($\sim\!40$\,Mpc distance) SGRB jet gave us an unprecedented opportunity to study its full angular structure. 

The other major result related to GW170817 is the first clear photometric and spectroscopic identification of a kilonova, i.e.~a UV/optical/IR transient powered by the radioactive decay of heavy r-process elements synthesized within the matter ejected by the merger process (e.g., \cite{Arcavi2017,Coulter2017,Pian2017,Smartt2017,Kasen2017}; see also \cite{Metzger2019LRR} and refs.~therein).
This confirmed that {\it BNS mergers produce a significant amount of elements heavier than iron, up to very large atomic mass numbers} ($A\!>\!140$).

\section{Open questions and ongoing research}
\label{open}

\noindent The discoveries connected with GW170817 certainly represent a breakthrough in the field, but a lot remains to be understood on both this event and BNS mergers in general. 
Part of our ignorance can be ascribed to current observational limits. For instance, much better constraints on the NS EOS will become available with the considerably higher sensitivity of third generation GW detectors \cite{Punturo2010,CE}, allowing also for a confident detection of the post-merger GW signal, while the merger rate, the formation scenarios of BNS systems, and the GW-based Hubble constant determination will strongly improve with the increasing number of detections. 
On the other hand, there are many other aspects for which the information encoded in the observed signals (in particular in the EM counterparts) could not be fully exploited due to the present limitations of theoretical models. 
This urgently calls for further development on the theory side and in particular in the context of BNS merger simulations in general relativity, which represents the leading approach to unravel the physical mechanisms at work when two NSs merge together. 

In the following, we discuss recent results of BNS merger simulations and the associated limitations, focussing the attention on the interpretation of the August 2017 event. In particular, we consider the two most important EM counterparts of this event, (i) the SGRB and its multiwavelength afterglows and (ii) the kilonova transient.

\subsection{SGRB central engines and GRB\,170817A}

\noindent Understanding the launching mechanism of a SGRB jet from a BNS merger and the nature of the remnant object acting as central engine is among the main driving motivations for the development of numerical relativity simulations of such mergers (e.g., \cite{Rezzolla2011,Kiuchi2014,Kawamura2016,Ruiz2016,Ciolfi2017,Ciolfi2019}).
The great progress of this type of simulations, especially over the last decade, allowed to reach important conclusions, even though the final solution of the SGRB puzzle is still ahead of us.

According to the most discussed scenario, a SGRB jet would be launched by a spinning BH surrounded by a massive ($\sim\!0.1\,M_\odot$) accretion disk, which is a likely outcome of a BNS merger.
Recent simulations \cite{Just2016,Perego2017a} showed that a jet powered by neutrino-antineutrino annihilation would not be powerful enough to explain the phenomenology of SGRBs, reinforcing the idea that SGRB jets should be instead magnetically driven.
Various GRMHD simulations explored the latter possibility (e.g., \cite{Rezzolla2011,Kiuchi2014,Kawamura2016,Ruiz2016}), confirming the formation of a low density funnel along the BH spin axis and finding indications of an emerging helical magnetic field structure that is favourable for accelerating an outflow. In addition, simulations reported in \cite{Ruiz2016} were the first to show the actual production of a magnetically-dominated mildly relativistic outflow and the authors argued that such an outflow could in principle reach terminal Lorentz factors compatible with a SGRB jet. 
While the results obtained so far do not provide the ultimate answer, current simulations suggest that the accreting BH scenario is a promising one (see, e.g., \cite{Ciolfi2018,Ciolfi2020b} for a more detailed discussion).
\begin{figure}[t!]
\includegraphics[width=0.5\textwidth]{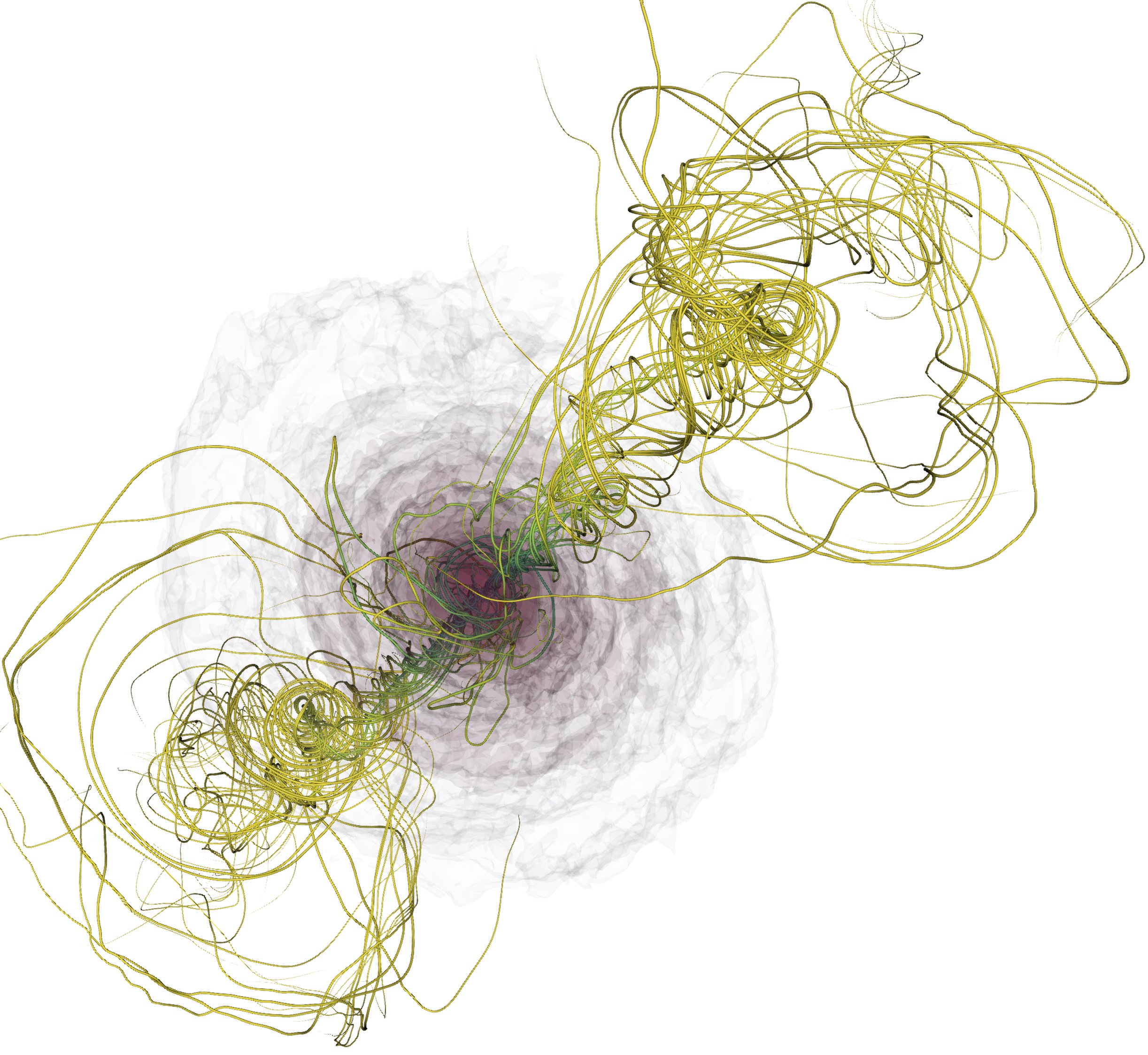}
\caption{Collimated helical magnetic field structure emerging along the spin axis of a long-lived BNS merger remnant (from a simulation presented in \cite{Ciolfi2020a}). 
Several semi-transperent isodensity surfaces are also shown for the highest rest-mass density region (with density increasing from grey to red).}
\label{fig2}
\end{figure}

The alternative scenario in which the central engine is a massive NS remnant is also investigated via GRMHD BNS merger simulations, although a systematic study started only recently \cite{Ciolfi2017,Ciolfi2019,Ciolfi2020a}. 
In this case, the higher level of baryon pollution along the spin axis may hamper the formation of an incipient jet. 
The longest (to date) simulations of this kind, recently presented in \cite{Ciolfi2020a}, showed for the first time that the NS differential rotation can still build up a helical magnetic field structure able to accelerate a collimated outflow (see Fig.~\ref{fig2}), although such an outcome is not ubiquitous.\footnote{ 
Note that a collimated outflow was also reported in studies where an ad-hoc dipolar field was superimposed by hand on a differentially rotating NS remnant (e.g., \cite{Shibata2011,Siegel2014,Moesta2020}).}
In addition, for the case under consideration, the properties of the collimated outflow (and in particular the very low terminal Lorentz factor) were found largely incompatible with a SGRB jet \cite{Ciolfi2020a}. 
This result reveals serious difficulties in powering a SGRB that might apply to massive NS remnants in general, thus pointing in favour of the alternative BH central engine. 
In order to confirm the above conclusion, however, a larger variety of physical conditions needs to be explored (e.g., by including neutrino radiation). 

For the case of GRB\,170817A, neither the observations nor current theoretical models can confidently exclude any of the two scenarios. Nevertheless, BNS merger simulations already provided valuable hints in favour of the accreting BH one \cite{Ruiz2016,Ciolfi2020a} and the continuous improvement of numerical codes and the degree of realism of their physical description might soon lead to a final solution.
 
Another important limiting factor for the interpretation of GRB\,170817A is represented by the considerable gap between the relatively small timescales and spatial scales probed by GRMHD merger simulations (up to order $\sim\!100$\,ms and $\sim\!1000$\,km) and those relevant for the propagation of an incipient jet through the baryon-polluted environment surrounding the merger site ($\gtrsim\!1$\,s and $\gtrsim\!10^5$\,km).
The ultimate angular structure and energetics of the escaping jet, which are directly related with the prompt and afterglow SGRB emission, are therefore very hard to associate with specific properties of the merging system and a specific launching mechanism. 
One of the most important challenges to be addressed in the near future will then be to obtain a self-consistent model able to describe the full evolution from the pre-merger stage up to the final escaping jet.

\subsection{Merger ejecta and the kilonova transient AT\,2017gfo}

\noindent During and after a BNS merger, a relatively large amount of material (up to $\sim\!0.1~M_\odot$) can be ejected, either via dynamical mechanisms associated with the merger process (tidally- and shock-driven ejecta) or via baryon-loaded winds launched by the (meta)stable massive NS remnant and/or by the accretion disk around the newly formed BH (if any).
Depending on the thermodynamical history and composition (in particular the electron fraction $Y_e$) of each fluid element within the ejecta, the r-process nucleosynthesis takes place producing a certain amount of heavy elements (i.e.~heavier than iron). Later on, the radioactive decay of these elements powers the thermal transient commonly referred to as a kilonova (e.g., \cite{Metzger2019LRR} and refs.~therein).

For a given ejecta component, the peak luminosity, peak time, and peak frequency (or temperature) of the corresponding kilonova are mainly determined by the ejecta mass, velocity, and opacity (e.g., \cite{Grossman2014}). While mass and velocity depend on the mass ejection mechanism, the opacity is directly related to the nucleosynthesis yields. In particular, high electron fractions ($Y_e\!\gtrsim\!0.25$) typically produce elements up to atomic mass numbers $A\!\lesssim\!140$, maintaining a relatively low opacity of $\sim\!0.1-1$\,cm$^2$/g, while more neutron rich ejecta ($Y_e\!\lesssim\!0.25$) allow for the production of elements up to $A\!>\!140$ (including, e.g., the group of lanthanides), which leads to much higher opacities of $\sim\!10$\,cm$^2$/g (e.g., \cite{Kasen2013,Tanaka2013}).

When applied to the kilonova of August 2017, the above picture reaveals that the observed transient (AT\,2017gfo) was generated by at least two distinct ejecta components (e.g., \cite{Kasen2017}),\footnote{Three-component models were also proposed (e.g., \cite{Perego2017b}).} one having mass $\approx\!1.5\!-\!2.5\!\times\!10^{-2}\,M_\odot$, velocity $\approx\!0.2\!-\!0.3\,c$, and a relatively low opacity of $\approx\!0.5$\,cm$^2$/g, leading to a ``blue'' kilonova peaking at $\sim\!1$\,day after merger, and the other having mass $\approx\!4\!-\!6\!\times\!10^{-2}\,M_\odot$, velocity $\approx\!0.1\,c$, and a much higher opacity of $\sim\!10$\,cm$^2$/g, leading to a ``red'' kilonova emerging on a timescale of $\sim\!1$\,week.
One of the current challenges is to identify the mass ejection mechanisms responsible for these components.
In such a quest, numerical relativity simulations of BNS mergers play a pivotal role. 

The ``red'' part of the 2017 kilonova is perhaps the easiest to accomodate (among the two). The very large mass and low velocity would exclude dynamical mass ejection and point to a baryon-loaded wind.
In particular, the mass expelled by the accretion disk around the BH (i.e. after the collapse of the NS remnant) appears to match the requirements, including a relatively high opacity or, equivalently, a low electron fraction for at least for part of the material (e.g., \cite{Siegel2018}).

The origin of the ``blue'' kilonova is more debated. Ejecta mass is rather high, but so is also the velocity ($v\!\gtrsim\!0.2\,c$). The former still poses doubts on a dynamical ejection, while the latter represents a potential problem for post-merger baryon-loaded winds.
The magnetically driven wind from the (meta)stable NS remnant offers a viable solution \cite{CiolfiKalinani2020}, thanks to the enhanced mass ejection and the simultaneous acceleration due to the magnetic field (as previously suggested, e.g., in \cite{Metzger2018}).
In this case, neutrino irradiation would also be fundamental to raise the $Y_e$ of the material, limit the r-process nucleosynthesis, and thus maintain a low opacity \cite{Metzger2018}.
We stress, however, that other viable scenarios exist (e.g., \cite{Kawaguchi2018,Nedora2019}).

Current kilonova models are still affected by several uncertainties on the microphysical parameters, on the radiation transport (which is treated with strong approximations), and on the mass ejection mechanisms. Nonetheless, we are witnessing a rapid theoretical and numerical progress that will keep guiding us towards a more solid interpretation of the observational data.

\section{Concluding remarks}
\label{concl}

\noindent The growing interest for BNS mergers over the last decades was recently boosted by the multimessenger observation of the August 2017 event. 
Among the numerous breakthrough results, this BNS merger provided fundamental confirmations of theoretical predictions, namely the association with SGRBs, already supported by indirect evidence but still unproven, and the production of heavy r-process elements and the related kilonova transients. 
This success on the theory side certainly strengthens the motivation for the development of models and in particular numerical simulations. 
At the same time, the case of GW170817 showed that the present and near future observations are likely to contain much more information than we are currently capable to exploit, making a further advancement in our ability to interpret the data more urgent than ever.

%

\bibliographystyle{apsrev4-1-noeprint}

%

\end{document}